\documentclass[11pt]{article}

\usepackage{amsmath}
\usepackage{mathtools}
\usepackage{etoolbox}
\usepackage{breqn}
\usepackage{url}
\usepackage{graphicx}
\usepackage{grffile}
\usepackage[left=1.5cm,right=2cm,top=2cm]{geometry}
\usepackage[font=small]{caption}
\usepackage{cite}
\usepackage{changes}
\usepackage{setspace}
\usepackage{booktabs}
\usepackage{multirow}
\usepackage{bm}
\usepackage{amssymb}
\usepackage{placeins}
\usepackage{lineno}

\makeatletter
\DeclareRobustCommand{\em}{%
  \@nomath\em \if b\expandafter\@car\f@series\@nil
  \normalfont \else \itshape \fi}
\makeatother

\newcommand{\commentout}[1]{}

\def\optionaltext#1{}

\def\ourtitle{Regulation of community functional composition across taxonomic variation by resource-consumer dynamics}

\begin{document}

\commentout{
\vspace{1.5in}
\large
\thispagestyle{empty}
\noindent{\bf \ourtitle}
\normalsize

\begin{flushleft}
Lee Worden$^{1}$
\end{flushleft}

\begin{flushleft}
$^1$Francis I. Proctor Foundation for Research in Ophthalmology, Box 0412, University of California, San Francisco, California 94143-0412
\end{flushleft}

\begin{flushleft}
\end{flushleft}

\begin{flushleft}
\today
\end{flushleft}

\vfill

\noindent Running head: Ecological regulation of traits

\newpage
}

\title{\ourtitle}

\author{Lee Worden}

\maketitle

\modulolinenumbers[2]

\begin{abstract}
High-throughput sequencing techniques such as
metagenomic and metatranscriptomic technologies allow cataloging
of functional characteristics of microbial community members as well
as their taxonomic identity.
Such studies have found that a community's composition
in terms of ecologically relevant functional traits or guilds can be
conserved more strictly across varying settings than
taxonomic composition is.
I use a standard ecological resource-consumer model to examine the dynamics
of traits relevant to resource consumption, and analyze determinants of
functional composition.
This model demonstrates that interaction with essential resources can regulate
the community-wide abundance of ecologically relevant traits,
keeping them at consistent levels
despite large changes in the abundances of the species housing those traits
in response to changes in the environment,
and across variation between communities in species composition.
Functional composition is shown to be able to track differences in
environmental conditions faithfully across differences in community
composition.
Mathematical conditions on consumers' vital rates and functional
responses sufficient to produce conservation of functional community
structure across taxonomic differences are presented.
\end{abstract}

\section{Introduction}

Microbes play a key role in every ecological community on earth,
and are crucial to the health of plants and animals
both as mutualists and as pathogens.
Understanding the ecological function and dynamics of microbes
is important to human health and to the health of the planet.
Because microbes exhibit short generation times, rapid evolution,
horizontal transmission of genes, and great diversity,
and can coexist in a massive number of partially isolated local communities,
the study of their communities can bring different questions
to the fore than are raised in the more common traditions
of ecological theory focused on plants and animals.
Newly available techniques of high-throughput genetic and
transcriptomic sequencing are making microbial community structure
visible in detail for the first time.

One pattern appearing in microbial communities,
in multiple very different settings,
is that communities composed very differently in terms of
species, genera, and even higher-level classifications of microbes
can have much more similar
structure when viewed in terms of the functional genes and genetic
pathways present in the communities
than when a catalog of taxa is constructed.
Additionally, environmental changes can induce consistent changes
in community-level abundance of relevant genes or pathways
while leaving others unchanged, in situations where such a pattern is
not readily visible in taxonomic data due to high taxonomic variability
across communities.
Metagenomic sequencing of samples
collected from a variety of ocean settings around the world shows
high taxonomic variability (even at the phylum level) with relatively
stable distribution of categories of functional genes
\cite{sunagawa_structure_2015},
and that
the environmental conditions predict the composition
of the community in terms of functional
groups better than in taxonomic structure,
suggesting that functional and taxonomic structure may constitute
roughly independent ``axes of variation'' in which functional structure
captures most of the variation predicted by environmental conditions
\cite{louca_high_2017}.
The same pattern of conserved functional community structure
across variation in taxonomic structure is seen in the human
microbiome
\cite{turnbaugh_core_2009,consortium_structure_2012,gosalbes_metagenomics_2012,gosalbes_metatranscriptomic_2011},
and in microbial communities assembled \emph{in vitro} on a single
nutrient resource \cite{goldford_emergent_2018}.
Convergence of functional community structure
with variation in species structure as a result of assembly history
is also seen in plant communities \cite{fukami_species_2005},
suggesting that explaining this pattern can have application beyond
microbial ecology.
A study of functional structure in \emph{in vitro} community assembly
\cite{goldford_emergent_2018} presents a mathematical model based
on the MacArthur consumer-resource dynamics model, which
numerically reproduces this pattern, but the model
is not analyzed.

Here I present a general class of consumer-resource models 
that describes
the community-wide abundances of functional traits
together with the abundances of species,
and analyze these models to explain how
regularity of functional structure can be an outcome
despite variability in species composition,
and when this outcome can occur in communities
governed by resource-consumer interactions.
I have used these models to construct a series of
simulation experiments applying this result
to functional community structure across variation
in enviromental conditions and in taxonomic community
structure.

First I tested a scenario in which functional
structure was preserved in a single community
across changes in species abundances as its environment changes.
Second I turned to the question of when multiple
communities converge to a common functional structure
despite differing taxonomic composition.
I present mathematical analysis of when this result occurs,
and then three model examples.
In one example, functional structure coincided with
high-level (genus or higher) taxonomic composition, and
community structure at that level was conserved across multiple
communities with different histories of assembly and
different species composition.
In the second, functional traits were shared across
taxa and co-occurring in diverse compositions within organisms,
so that functional structure was not reflected at a higher
taxonomic level, and conservation of functional structure
was achieved by a complex balance of functionally overlapping species.
Third is a simulated controlled experiment in which
selected traits were upregulated and downregulated by manipulation
of the environment while other traits were unaffected,
in a community model similar to the second, above.

\section{Trait abundances in consumer-resource models}

A standard model framework for resource-consumer dynamics
is widely used and well understood, particularly given a finite number of
distinct species without spatial patchiness
\cite{macarthur_competition_1964,levin_community_1970,tilman_resource_1982}.

Resource abundances are increased by supply from outside the
model community, and decreased by uptake by consumer species,
and species abundances are increased by reproduction at a rate
that depends on resource consumption, and decreased by
fixed per-capita mortality.
For example, one such model has this form:
\begin{dgroup*}
\begin{dmath*}
 \frac{dX_i}{dt} = \sum_j c_{ij} r_{ij} R_j X_i - m_i X_i \condition{$i=1,\ldots,n_s$}
\end{dmath*}
\begin{dmath*}
 \frac{dR_j}{dt} = s_j - \sum_i r_{ij} R_j X_i \condition{$j=1,\ldots,n_r$}
\end{dmath*},
\end{dgroup*}
where $X_i$ is the abundance of consumer species $i$ and
$R_j$ is the abundance of resource $j$, while
$r_{ij}$ is the consumption rate of resource $j$ by consumer $i$,
$c_{ij}$ is a conversion rate of resource $j$ into reproductive fitness of $i$,
$m_i$ is the per-capita mortality rate of consumer $i$,
and $s_j$ is the rate of supply of resource $j$.

To analyze the behavior of functional traits and
genes across the community, it is necessary to include
a definition of trait abundance in the model.
Let us assume that a species that consumes a given resource
has a trait of consumption of that resource.
Thus given $n_r$ resources I define the corresponding $n_r$ traits,
one for each, which each consumer may possess or not:
let $A_{ij}$ be one if consumer $i$
has trait $j$ and zero if not, and let $f_{ij}(\mathbf{R})$,
the functional response, or uptake rate of resource $j$ by consumer $i$, be
a continuous, nondecreasing, nonnegative
function of the vector $\mathbf{R}$ of resource abundances. 
Trait assignments $A_{ij}$ that take on a greater range of
nonnegative values may also be of interest, for future research.

Including this description of trait possession,
a consumer-resource dynamics model has the form
\begin{dgroup*}
\begin{dmath*}
\frac{dX_i}{dt} = \sum_j A_{ij} c_{ij} f_{ij}(\mathbf{R}) X_i - m_i X_i \condition{$i=1,\ldots,n_s$}
\end{dmath*}
\begin{dmath*}
\frac{dR_j}{dt} = s_j - \sum_i A_{ij} f_{ij}(\mathbf{R}) X_i \condition{$j=1,\ldots,n_r$}
\end{dmath*}.
\end{dgroup*}

A type~I functional response has the linear form
\begin{dgroup*}
\begin{dmath}\label{eqn:typei}
f_{ij}(\mathbf{R}) = r_{ij} R_j
\end{dmath},
\end{dgroup*}
and a type~II functional response (e.g.~\cite{holling1959some})
can take at least two forms:
\begin{dgroup*}
\begin{dmath}\label{eqn:typeii}
f_{ij}(\mathbf{R}) = \frac{r_{ij} R_j}{1 + h_{ij} R_j}
\end{dmath},
\end{dgroup*}
or
\begin{dgroup*}
\begin{dmath}\label{eqn:typeiicomplex}
f_{ij}(\mathbf{R}) = \frac{r_{ij} R_j}{1 + h_{ij}\sum_k A_{ik} R_k}
\end{dmath},
\end{dgroup*}
depending on whether saturation occurs independently for each
trait a consumer possesses,
with $h_{ij}$ as a constant describing how quickly resource consumption
saturates in response to its availability.
The functional response may also be a type~III response
\cite{holling1959components},
which can be described by a variety of mathematical forms.
In the example model systems I present below,
I use the type~I and the
first of the above two type~II functional response forms.

There are at least two measures of abundance that can be used,
motivated by forms of next-generation sequencing in widespread use.

Using a measure of \emph{possession} of genes,
as seen in metagenomic sequencing processes based on DNA
sequences,
the community-wide abundance of trait $j$ is defined as
the total value of $A_{ij}$ over all consumers:
\begin{dgroup*}
\begin{dmath*}
T_j = \sum_i A_{ij} X_i
\end{dmath*}.
\end{dgroup*}

A measure of \emph{expression} of traits,
more like the data reported by metatranscriptomic sequencing
processes such as RNA-Seq,
describes not the presence of genetic sequences but
the rates at which their functions are actively used:
\begin{dgroup*}
\begin{dmath*}
E_j = \sum_i A_{ij} X_i f_{ij}(\mathbf{R})
\end{dmath*}.
\end{dgroup*}

This paper analyzes conditions under which trait
abundances $T_j$ and $E_j$ remain unchanged or nearly so while
species abundances $X_i$ vary,
and when species composition, in the sense of the presence and absence
of specific species in a community, varies across communities.
I present conditions for conservation of both measures of traits
across environmental conditions and community structures,
and examples in which the abundances of genetic material
$T_j$ are conserved, the more stringent case.

\subsection{Analysis of consumer-resource models}

Given the above form of model,
the behavior of these models is well understood
\cite{macarthur_competition_1964,levin_community_1970,tilman_resource_1982}.
When the community consists of a single consumer species
dependent on a single limiting resource,
the population size grows until its increasing
resource consumption lowers the resource abundance to a level at which
the consumer's reproduction and mortality rates balance.
In this way, the resource abundance is regulated by the
consumer: the abundance of the resource at equilibrium is a quantity
determined by those organisms' processes of reproduction and
mortality.

The population brings its limiting resource to the same
equilibrium level, conventionally known as $R^*$
\cite{tilman_resource_1982},
regardless of whether the flow of the resource into the community
is small or large.
If there is a large inflow, the population size grows until it
is consuming the resource at an equally high rate,
drawing the resource abundance down to the required level.
If inflow is small, population size becomes as small as needed
to balance the flows.
In this way, the size of the population is determined by the
resource supply rate,
but the abundance of the resource is not.

When there are multiple species and multiple resources,
for each species there are certain combinations of resource abundances
that balance its birth and death rates.
With $n_s$ species and $n_r$ resources, these equilibrium conditions take
the form of $n_s$ equations, one for each population $X_i$,
each in $n_r$ unknowns $R_j$:
\begin{dgroup*}
\begin{dmath}\label{eq:rstar-equilibrium}
 \frac{dX_i}{dt} = \sum_j A_{ij} c_{ij} f_{ij}(\mathbf{R}^*) X^*_i - m_i X^*_i \hiderel{=} 0
 \condition{for each $i$}
\end{dmath},
\end{dgroup*}
where $\mathbf{R^*}$ is the vector
$(R^*_1,\ldots,R^*_{n_r})$ of equilibrium values of the resource
abundances.
Each of these equations, one for each $i$, can in principle be solved for
the set of values of $R^*_1$ through $R^*_{n_r}$ that satisfy
this condition.
Note that these solutions are not affected by the
population sizes $X^*_i$ as long as the population sizes are nonzero.
The solution set of the $i$th equation describes the set of values of the
$n_r$ resources at which net growth of species $i$ is zero.
The solution of all these equations simultaneously is the
set of resource abundances at which all species' growth
is at equilibrium.
This is why $n_r$ resources can support at most $n_r$
coexisting populations in these models under most conditions:
because outside of special cases,
no more than $n_r$ equations can be solved
for $n_r$ variables simultaneously
\cite{macarthur_competition_1964,levin_community_1970}.
The equilibrium resource abundances $R^*_j$, taken together,
are the solution of that system of equations.
Thus the combination of all $n_r$ resource abundances at equilibrium
is determined by the requirements of all the consumer populations
combined.
Note that they are independent of the resource supply rates as well
as of the consumer population sizes.

That balance of resources is enforced
by the sizes of consumer populations: 
if resources increase above the levels that produce consumer equilibrium,
consumer populations grow, drawing resources at increased rates,
and the opposite if resource levels drop,
until the resources are returned to the required levels and
supply rates are matched by the rates of consumption.
Equilibrium levels of consumers are determined not by the
above equilibrium equations, but by the model's other set of
equilibrium conditions:
\begin{dgroup*}
\begin{dmath} \label{eq:xstar-equilibrium}
 \frac{dR_j}{dt} = s_j - \sum_i A_{ij} f_{ij}( \mathbf{R} ) X_i \hiderel{=} 0
\end{dmath}.
\end{dgroup*}
At equilibrium, the consumer population sizes must be
whatever values $X^*_i$ are required to make the overall
uptake rate of each
resource $j$ described by this equation equal to the supply rate $s_j$,
when the resources are at the levels $R^*$ implied
by the earlier equilibrium conditions (\ref{eq:rstar-equilibrium}).
Thus the consumer population sizes, all taken together,
are determined by the
supply rates of all the relevant resources taken together,
given the equilibrium resource levels,
in such a way that resource inflow and outflow rates are balanced.
When each resource is controlled by multiple consumers all of whom
use multiple resources, each consumer abundance is determined by
all the resource supplies in balance with the other consumers in
ways that may be difficult to predict or explain.

In summary,
there is a duality of causal relationships between the two
players in this system, resources and consumers,
in which resource levels are determined by the consumers' physiology
(\ref{eq:rstar-equilibrium}),
and consumer levels are determined
in a complex interlocking way by the resource supply rates
given the above resource levels
(\ref{eq:xstar-equilibrium}).

\subsection{Conditions for conservation of trait abundances across
differences in community composition}

Given an arbitrary assemblage of resource consumer species,
described by some unconstrained assignment of values to
the functions $f_{ij}$, mortality rates $m_i$, and
conversion factors $c_{ij}$, without knowledge of
those values nothing can be concluded about
the abundances of species, resources, and traits
that will be observed in the long term.

However, under certain constraints on the relationships between
these values, it can be shown that trait abundances at equilibrium,
given that enough consumer species coexist at equilibrium,
are determined only by the resources' supply rates
without dependence on the consumers' characteristics.

I have derived conditions for
simple dependence of trait abundances on their resources'
supply rates
in the appendix (\ref{app:analysis}),
and I summarize them here.

\textbf{Condition for conservation of rates of trait expression,
$\mathbf{E}$.}
The community-wide rate of expression of a trait,
labeled $E_j$ above, is determined by the
supply rate of its resource in all model communities of the
above form, provided that the community is consuming
all resources at equilibrium.
This result is simply because $E_j$ is tied to
the rate of resource uptake, which must match the supply rate
of the resource at equilibrium.

The community-wide abundance of possession of a trait ($T_j$) is
not tied to supply rates in all cases, but conditions exist
under which these abundances are directly predicted by supply
rates independent of species abundances.

\textbf{Simple condition for conservation of trait
abundances, $\mathbf{T}$.}
A condition for conservation of trait abundances $T_j$
is that there are constant numbers $k_j$, one for each resource $j$,
for which
\begin{dgroup*}
\begin{dmath}\label{eqn:diagonalcondition}
  \sum_j c_{ij} A_{ij} / k_j = m_i
\condition{for each species $i$.}
\end{dmath}
\end{dgroup*}
If that condition is met, and the response functions $f_{ij}()$
are defined in such a way that there is a set of resource levels
$R_j$ that can satisfy the constraint 
$ f_{ij}(\mathbf{R}) = 1/k_j $
for each $i$ and $j$ for which $A_{ij}>0$, at the same time,
then those resource levels describe an equilibrium for
each community structure, at which
community-wide trait abundance $T_j$ will be
held fixed at a level that depends only on the supply rate $s_j$,
even though community structure and species abundances may vary.

This condition can be explained by recognizing that the resource
uptake rates $f$ are a scaling factor between the raw trait abundances
$T$ and the trait expression rates $E$: since the expression rates are
in a fixed relation to supply rates across communities, for the
trait abundances to be fixed in that way as well, the ratio between
the two, which is $f_{ij}$, must be fixed.

\textbf{General condition for conservation of trait abundances,
$\mathbf{T}$.}
Condition (\ref{eqn:diagonalcondition}),
in which each trait abundance $T^*_j$ depends only on 
the supply rate of the one corresponding resource supply rate $s_j$,
is a special case of a more general case
in which the full
vector of trait abundances is determined by the full
vector of resource supply rates,
the condition for which is the more abstract one
that constants $K_{jk}$ exist for which
\begin{dgroup*}
\begin{dmath*}
  \sum_k K_{jk} A_{ik} f_{ik}(\mathbf{R}) = A_{ij}
\condition{for each $i$ and $j$,}
\end{dmath*}
\intertext{and}
\begin{dmath*}
  \sum_j c_{ij} A_{ij} f_{ij}(\mathbf{R}) = m_i
\condition{for each $i$,}
\end{dmath*}
\end{dgroup*}
simultaneously, for at least one value of $\mathbf{R}$.

\textbf{Approximate conservation of trait structure.}
If either of these relations is not exactly but very nearly satisfied,
then the model can almost exactly conserve the trait abundances.
See (\ref{app:analysis}) for more formal discussion of this point
and derivation of the above conditions.

\textbf{Simple construction of example models.}
Examples in this paper, below, are constructed using the simple condition
(\ref{eqn:diagonalcondition}) for conservation of trait abundances,
by assigning all mortality rates equal to a constant value $m$,
conversion rates $c_{ij}$ equal to a constant $c$,
with a fixed number $p$ of traits assigned to each consumer.
This satisfies (\ref{eqn:diagonalcondition}) with $k_j = m/cp$ for each $j$.

Under these conditions, given that there is sufficient diversity
within a community to fix resource abundances at the required equilibrium
point, 
each trait $T_j$ will have equilibrium abundance
$T^*_j=s_j/k_j$,
independent of the trait assigments $A_{ij}$
and species abundances $X^*_i$.
Species abundances are implied by the definition
$T^*_j = \sum_i A_{ij} X^*_i$,
and vary with the specifics of the community structure.

\section{Example: Complex regulation of functional structure within a community}

The above analysis implies that
abundance of each of a palette of traits can be regulated
by the availability of the one resource associated with that trait,
even though every organism in the community possesses multiple such traits.
I observed the regulation of community-wide trait abundances
within a single community
using a model of four resources and four consumers.
For each resource I defined a trait corresponding to consumption
of that resource, which was shared by multiple consumer
populations.
The first three resources were supplied at a constant rate,
while the fourth resource was supplied at a rate that changed at
discrete times.
Species and trait abundances shifted in response
to the changing supply of the fourth resource.



In this model, with type I functional responses as in (\ref{eqn:typei}),
the trait assignments $A_{ij}$ were constructed such that
each species consumed a different three of the four
resources (Fig.~\ref{fig:fixed-traits-results}A).
The resource supply rate constants $s_j$ were held constant
for resources 1, 2, and 3, while $s_4$ was piecewise constant,
changing between three different values at discrete moments
(Figure~\ref{fig:fixed-traits-results}B).

The abundances of the four consumer species making up the model
community came to equilibrium when their habitat was unchanging,
but when the supply rate of resource 4 changed, they
all shifted to different equilibrium levels
(Figure~\ref{fig:fixed-traits-results}C).
However, despite these complex shifts in all the consumer species'
abundances,
the community-wide abundances of the traits of consumption of the first
three resources were conserved at equilibrium across these changes in
community structure, aside from brief transient adjustments
(Figure~\ref{fig:fixed-traits-results}D).
Of the four traits modeled, only the fourth changed in equilibrium abundance
in response to the changing resource supply.

The community was able to regulate the community-wide abundance of
the trait involved in consumption of the fourth resource independent
of the other three consumption traits, despite that fact that
multiple of the four traits coexisted in every organism in the community.

\newcommand{\figwlabel}[3][width=0.9\textwidth]{
\begin{minipage}[t]{0.02\textwidth}
\vspace{0pt}
\textbf{#2}
\end{minipage}
\begin{minipage}[t]{0.46\textwidth}
\vspace{0pt}
\includegraphics[#1]{#3}
\end{minipage}
}

\begin{figure}[h!]
\centering
\begin{tabular}{ll}
\figwlabel{A.}{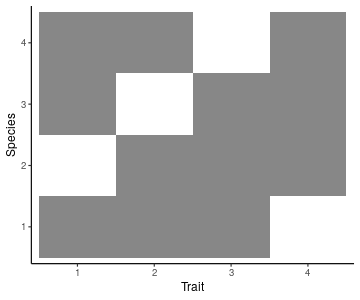}
&
\figwlabel{B.}{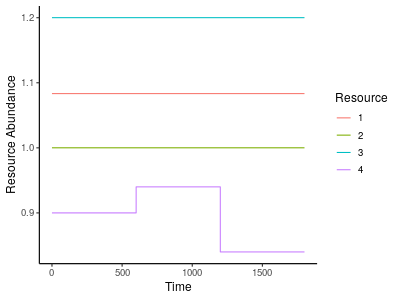}
\\
\figwlabel{C.}{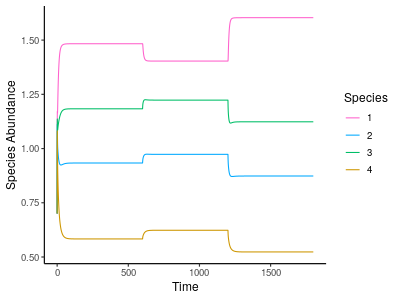}
&
\figwlabel[width=0.84\textwidth]{D.}{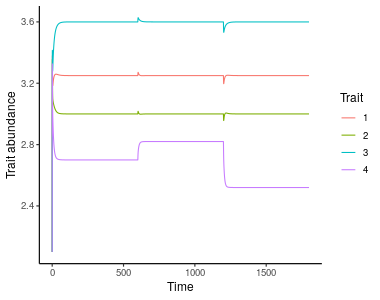}
\end{tabular}
\caption{
\label{fig:fixed-traits-results}
{\bf Regulation of functional structure within a community.}
A community model that satisfies the condition (\ref{eqn:diagonalcondition})
maintains trait abundances fixed through changing environmental conditions
by rebalancing all consumer population sizes.
{\bf A. Assignment of traits to species} (dark=present, white=absent);
{\bf B. Supply rates of resources}, with resources 1 through 3 supplied
at constant rate, supply of resource 4 changing at discrete times.
{\bf C. Species abundances} all vary with changes in supply of
resource 4, while
{\bf D. Whole-community trait abundances} 1 through 3 are constant
apart from transient fluctuations
with only trait 4 changing in response to changing supply of resource 4.
}
\end{figure}

This model community achieved equilibrium by bringing trait abundances
to the needed levels after each change in community structure, even though
the sizes of the four populations embodying those trait abundances were all
different after each change.
The population sizes were all altered in just the way necessary
to adjust the total abundance of the trait of consumption of resource 4
to match the changing supply rate of resource 4
and leave the other three unaltered.



\section{Example: Conservation of functional groups across differences in species composition}

One possible way in which communities may have a functional regularity
that is not captured at the species level is that species
may be interchangeable within guilds or groups of 
functionally equivalent species, where total counts in
a group are conserved while species composition is not.
The species in a group may be members of a family or phylum, or
may be unrelated but perform similar functions.
I constructed a model involving multiple guilds,
in which members of each guild shared a
functional trait of consumption of a guild-defining resource
and varied in other traits.
Communities were assembled by drawing species
from a common pool of species.

Consumer species were grouped into three guilds,
each guild defined by consumption of a guild-specific resource,
with each species belonging to exactly one guild.
Each species also consumed three other resources,
assigned randomly from a common pool of five resources without regard
to guild membership.
Consumers' functional response to resources was type II (\ref{eqn:typeii}).
For parsimony, resource supply rates were set equal at a
numerical value of $3/2$,
and the saturation parameters $h_{ij}$,
ideal uptake rates $r_{ij}$, conversion factors $c_{ij}$,
and mortality rates $m_i$ were all set to 1.

\begin{figure}
\centering
\begin{tabular}{ll}
\figwlabel[height=0.6\textwidth]{A.}{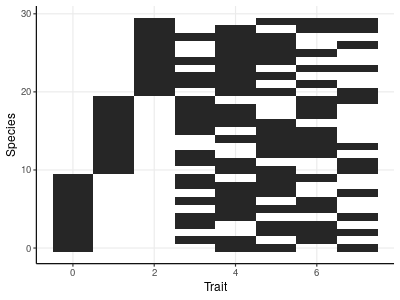} &
\figwlabel[height=0.6\textwidth]{B.}{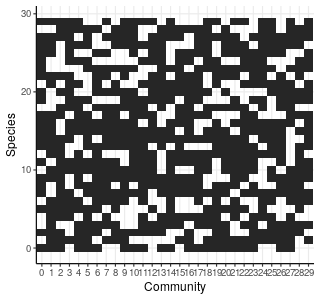} \\
\figwlabel[height=0.6\textwidth]{C.}{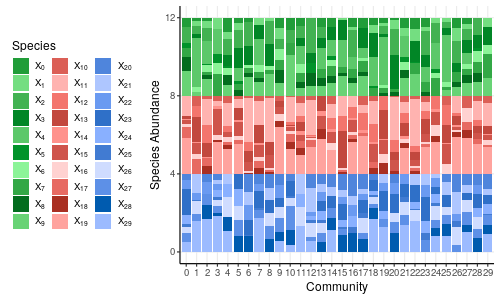} &
\figwlabel[height=0.6\textwidth]{D.}{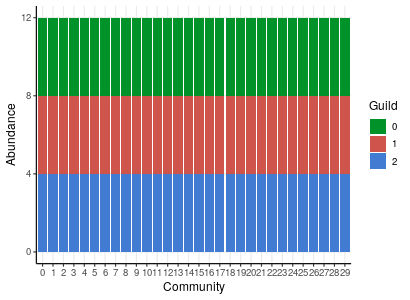}
\end{tabular}
\caption{ \label{fig:guild}
\textbf{Conservation of functional groups across differences in species composition.}
Overall abundances of each of three ``guilds'' of consumers of different resources
are held fixed across communities assembled randomly from varying species of each
guild.
(\textbf{A.}) \textbf{Assignment of traits to species}
in guild model.
Guild membership is defined by the first three traits,
corresponding to consumption of ``guild-defining resources,''
while the other five traits are guild-independent traits
that distinguish species from one another.
(\textbf{B.}) \textbf{Assignment of species to communities}.
Some species assigned to communities may not survive beyond an initial
transient as community comes to equilibrium.
(\textbf{C.}) \textbf{Species abundances}
(color coded by guild membership)
and (\textbf{D.}) \textbf{overall abundances of guilds}
at equilibrium, by community,
in guild model.
}
\end{figure}

Thirty species were constructed, ten in each guild,
by assigning non-guild traits at random
conditional on the species-trait assignment matrix having the
maximum possible rank%
\footnote{One fewer than the number of traits, or seven,
is the highest rank this matrix can attain, given the requirement
that guild traits sum to one and all traits sum to four for every species.},
and thirty communities were constructed by randomly assigning twenty-one
species to each, conditional on each community's ${A}$ matrix
having maximal rank
(Figure~\ref{fig:guild}A and B).

The dynamics of these model communities was evaluated,
starting from initial conditions at which all species and resource
abundances were $1.0$ in their respective units, for 200 time steps.
Total abundances in each guild at the end of that time were
plotted for comparison across communities.

At the end of that process,
species composition varied across communities (Figure~\ref{fig:guild}C),
but the overall abundance of each guild was uniform across
the model communities (Figure~\ref{fig:guild}D).

\section{Example: Conservation of overlapping functional traits across differences
in species composition}
\label{sec:overlapping}

While in the above model,
each species belonged to a single functional guild,
I constructed a second model in which each consumer
possessed multiple functional traits that were shared
by other consumer species in varying combinations, so that
regulation of trait abundances required a complex balancing of
all the overlapping species.

\begin{figure}
\centering
\begin{tabular}{ll}
\figwlabel[height=0.6\textwidth]{A.}{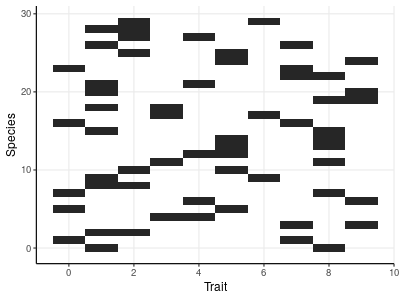} &
\figwlabel[height=0.6\textwidth]{B.}{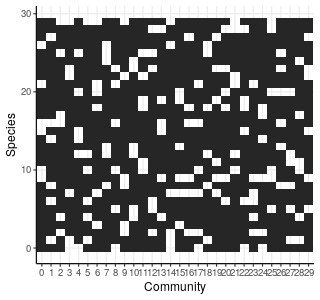} \\
\figwlabel[height=0.6\textwidth]{C.}{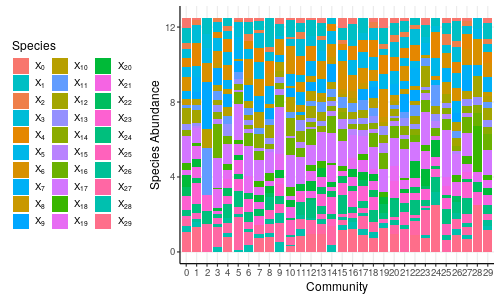} &
\figwlabel[height=0.6\textwidth]{D.}{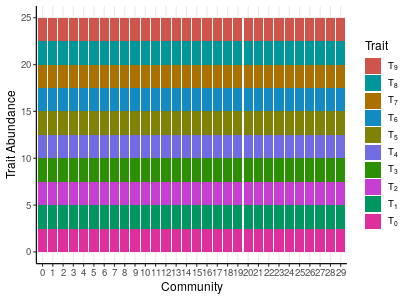}
\end{tabular}
\caption{ \label{fig:rc}
\textbf{Conservation of overlapping functional traits across differences
in species composition.}
When all consumption traits are randomly assorted across consumers,
overall trait abundances are equal, as predicted by equal resource
supply rates, independent of consumer species presence/absence or
abundances.
(\textbf{A.}) \textbf{Assignment of traits to species},
(\textbf{B.}) \textbf{assignment of species to communities},
(\textbf{C.}) \textbf{species abundances}
at equilibrium, by community,
and (\textbf{D.}) \textbf{trait abundances}
at equilibrium, by community,
in overlapping-traits model.
Some species assigned to communities may not survive beyond an initial
transient as community comes to equilibrium.
}
\end{figure}

The model was the same as above with the difference that
rather than assigning species to guilds characterized
by special traits, each species was assigned two of the ten
resource consumption traits at random
(Figure~\ref{fig:rc}A).
As above, I constructed 30 communities
by assigning 24 species to each,
chosen at random from a common pool of 30 candidate species,
conditional on full rank
(Figure~\ref{fig:rc}B).
Numerical parameters and functional responses were as in the above guild model,
except that here $m_i=6.4$, $c_{ij}=4$, and $s_i=2$ for all $i$ and $j$.
I recorded functional and taxonomic abundances after 200 time steps
from initial conditions of uniform resource and species abundances
of $1.0$.
At the end of that time I found that
species abundances varied widely from community to community,
but trait abundances were uniform across communities
(Figure~\ref{fig:rc}).


\section{Example: Coexistence of conserved and variable traits}

Where the above model results explored cases in
which functional community structure was the same across
communities due to an underlying equality in conditions,
here I look at how differences in conditions can be
reflected by predictable differences in functional structure.
The model I present here was constructed in the same way
as in Section~\ref{sec:overlapping},
with the difference that
15 communities were constructed and then evaluated
subject to two different environments,
labeled control and treatment.
All parameters were set as above in the control arm of
the experiment,
while in the treatment arm the resources were partitioned
into three classes, one whose supply rates were unchanged at 2.0,
one in which supply rates were elevated to 2.8,
and one which which they were reduced to 1.2.

\begin{figure}
\centering
\begin{tabular}{ll}
\figwlabel[height=0.6\textwidth]{A.}{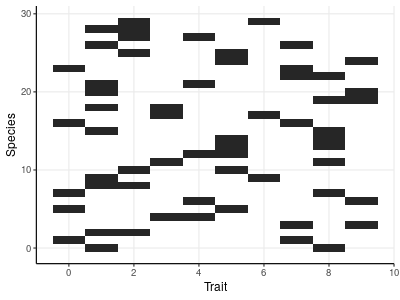} &
\figwlabel[height=0.6\textwidth]{B.}{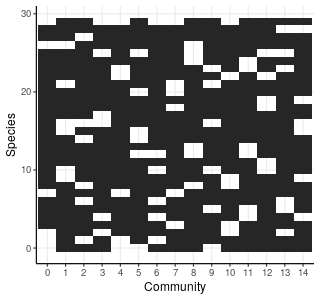} \\
\figwlabel[height=0.6\textwidth]{C.}{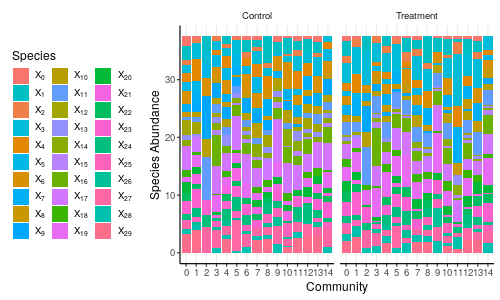} &
\figwlabel[height=0.6\textwidth]{D.}{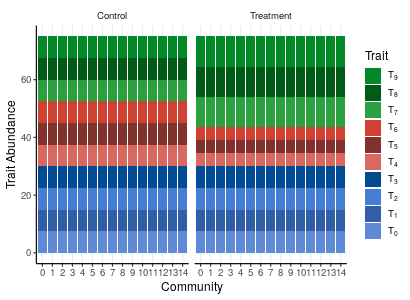}
\end{tabular}
\caption{ \label{fig:difference}
\textbf{Coexistence of conserved and variable traits in simulated
experimental conditions.}
Randomly assembled model communities are evaluated in ``control'' conditions
of equal resource supply rates, and ``treatment'' conditions with
altered supply rates.
Trait abundances track supply rates across differences in
community composition and across arms of the experiment.
(\textbf{A.}) \textbf{Assignment of traits to species},
(\textbf{B.}) \textbf{assignment of species to communities},
(\textbf{C.}) \textbf{Species abundances}
at equilibrium, by community and treatment arm,
and (\textbf{D.}) \textbf{trait abundances}
at equilibrium, by community and treatment arm,
in simulated experiment model.
Each community is simulated under both treatment and control conditions.
Some species assigned to communities may not survive beyond an initial
transient as community comes to equilibrium.
}
\end{figure}

Trait abundances at equilibrium (Figure~\ref{fig:difference}D)
clearly distinguished treated from
control communities, and treated from unmodified resources.
The traits associated with fixed-supply resources behaved like a
``core functional structure'' across these communities,
while the traits associated with treated resources were variable
in their abundances in accordance with the variation in resource supply.
Species abundances varied between communities in both control and treatment groups
(Figure~\ref{fig:difference}C),
and did not provide a visually apparent indicator of group membership.

\section{Discussion}

The above analysis has demonstrated, in a broad class of widely used
models of consumer-resource ecological community dynamics,
conditions on consumer physiology under which the community-wide
abundance of traits, pathways, or genes involved in resource
usage can be predicted by resource availability, independent of
the taxonomic makeup of the community and the abundances of
the taxa it includes.

In these model examples, and in general,
the total rate of uptake of a given resource
must balance the resource's net rate of supply.
Therefore if the supply does not change, and the
community continues to consume that resource,
the uptake rate must
return to the same level, after a possible transient fluctuation,
after a perturbation in the community,
and two communities comprised of different species
but encountering the same rates of supply
must necessarily manifest the same uptake rates.

In the models analyzed here,
this matching of outflow to inflow can cause
the community-wide rate of expression of traits associated with
resource consumption to be conserved.
If expression of a given trait or pathway is directly
related to the rate of consumption of a resource,
it is natural that the rate of trait expression
should be predicted by the rate of resource supply
because of the relation between supply rate and
overall uptake rate.
Even though those traits may be distributed across
multiple consumer species, each involved with multiple resources,
a central result of these models is that species abundances
are driven by resource availability in such a way as to regulate
all the resources simultaneously \cite{levin_community_1970},
even if that balance requires a complex adjustment of all the species
in the community.
In this way, community-wide rates of trait expression can be
matched to resource supplies even though all species abundances
may vary irregularly in whatever ways are needed to make uptake
rates balance supply rates.

Such a concept of trait expression (the quantity $E_j$ defined
above) is likely more appropriate to
a metatranscriptomic description of a community, as implemented by
high-throughput sequencing techniques such as RNA-Seq \cite{wang2009rna},
than to a metagenomic description.
A metagenomic description generated by measuring abundances of
DNA sequences in cells may be better described as a measure of
trait or gene prevalence, in the sense of abundance of
organisms possessing the genes,
such as the quantity $T_j$ used here.
This quantity can also be tied to resource supply rates
as trait expression can, but the relation is not as universal
and requires more conditions.
In summary, organisms that possess a genetic pathway
may use it at varying rates depending on the availability of
the resources involved, and on whether conditions are more
favorable to the use of other pathways.
In terms of the model dynamics, the uptake rate depends
both on the abundance of organisms possessing the relevant trait
and on the availability of the various resources relevant to
those organisms.
The conditions for conservation of the overall prevalence of such
genes across the community include regulation of resource 
uptake rates \emph{per capita} to common levels across communities.
In many if not all conditions, this likely requires control of
resource abundances to common levels across communities
(the vector $\mathbf{R}^*$).
Under those conditions, a fixed relation between trait expression
and trait prevalence is maintained, allowing both to be conserved
across differences in consumer species.

The examples in this article have demonstrated this more stringent
condition of regulation of trait prevalences, for illustration purposes.
They showed a series of different results involving regulation
of community functional structure, as defined by trait prevalences,
that can be manifest by this effect.

The first example demonstrated that in a single model community,
a temporal fluctuation in one resource supply rate
can induce a coordinated shift in all of the species abundances,
though its effect on the trait abundances is restricted to the one trait,
leaving the community's functional structure otherwise unchanged.

Second, in a model of guilds of consumers specialized on different
resources, that is, each characterized by a different resource-consumption
trait, the overall size of each guild was shown to be predicted
directly by resource supply, across multiple community structures,
while the species composition of each guild varied widely across
communities.
This follows from the definition of guilds which makes
their sizes effectively identical to trait prevalences.

Next was a model in which consumption traits were not partitioned into
disjoint guilds, but shared in overlapping ways
by consumers of multiple resources.
In multiple model communities differing widely in species composition,
the community-wide abundance of consumption traits was nonetheless
seen to be uniform across communities when resource supplies were
uniform.

Finally, differences in resource supply in a model controlled experiment
were shown to produce regular, predictable differences in trait prevalences
across model communities while core functional traits corresponding
to unaltered resources were held fixed, at the same time that
all species abundances varied across communities and treatment groups in
apparently irregular ways.

All the above examples demonstrated conditions in which functional
structure of a community is exactly determined by its environment,
independent of its taxonomic composition.
The theory that predicts these outcomes also predicts that
functional structure will be approximately conserved across community
structures if the conditions are nearly enough met,
an important consideration when attempting to apply such results
to the imprecise world of biology outside of models.

These results demonstrate a mechanism by which functional structure can be
predicted directly from environmental conditions in a simple case,
bypassing the complexities of taxonomic variation.
It should be read not as a faithful model to be applied directly
to communities in the lab or field, but as a step toward
a fuller theory to describe them.
This paper offers a proof of concept that
conservation of trait abundances can be explained
by known models of community dynamics,
and that functional observations of communities can
describe and predict their behavior more parsimoniously
than taxonomic observations.

Interestingly, these results turn out to be
insensitive to the range of different types of
functional response curves that can be manifested by resource consumers.
Instead, they require a condition of uniformity of response
across consumer taxa. The different consumers' responses to
resource availability must satisfy a somewhat opaque
consistency condition, to allow the
total presence of resource consumption traits to be held
in a consistent relation to resource supply at the same time that
the rates of trait expression are as well.
In the example model results presented above,
this is achieved by assuming the functional
response to each resource is the same among all consumers that
use it, which is an especially simple way to satisfy the condition,
regardless of the type of functional response.
Note that the result does not require that all species included
in community assembly 
meet these conditions, but only
that they be met by the consumers that are present in the community
at equilibrium.

This work has multiple limitations.
It does not apply directly to communities whose dynamics are
shaped by interactions other than resource competition,
for example bacteria-phage interactions, direct competition
or facilitation between microbes, or host-guest interactions
such as host immunity.
While similar results may hold in these cases, they require
expanded models to investigate them.
The assumptions made here about the close mapping between
trait expression and uptake rates are likely not satisfied
in many cases, and should
be unpacked to allow a fuller treatment of the subject.
Spatial heterogeneity alters the behavior of consumer-resource
models and must be studied separately,
and can open up additional interesting questions such
as the response of communities to spatially variable
resource supply.
Endogenous taxonomic heterogeneity driven by local dispersal
may not imply comparable functional heterogeneity if underlying
abiotic conditions are homogeneous.
The analysis of equilibrium community structures also
likely does not apply to many communities, and it may
be worthwhile to expand the analysis to describe slow dynamics of
community structure in conditions of constant immigration,
seasonal or other temporal variability in the environment,
or evolutionary change in which equilibrium is not attained.

It is not obvious whether the conditions presented here for
compatibility of vital rates across taxa to make trait
abundances behave regularly are realistic for microbes.
Having established that these are the necessary conditions
in these models, if they are not considered believable,
then this work serves to illuminate the questions that must
be answered about how microbial communities diverge from
these models, and how else their observed functional regularities
can be explained.
One avenue might be to investigate whether $R^*$ competition
under conditions of high diversity
can reduce a community without the closely matched $R^*$ conditions
described here to a subcommunity in which such a condition is
roughly though not precisely met.

These results suggest a number of further questions
to be investigated, such as the
impact of more complex mappings between genetic pathways
and resource uptake dynamics, and the
dynamics of functional community structure
in the presence of mechanisms such as
direct microbe-microbe interactions
or host immunity.
The dynamics of traits involved in functions other than
resource consumption
is left to be studied,
such as for example 
drug resistance or dispersal ability,
as is the impact of evolutionary dynamics,
including horizontal transfer, on the
dynamics of functional composition.
It may be productive to investigate whether a community's
need to regulate its functional composition in certain
ways can lead to selection for certain kinds of genetic
robustness, dispersal, horizontal transfer, or other
characteristics.
It would be of interest to study whether conditions
selecting for sharing of traits across taxa
can be distinguished from those selecting for
specialization by taxon.
The present study is offered as an initial investigation,
presenting an existence proof
of the ability of a community to regulate its
functional composition independent of its
taxonomic makeup, in hope it will open doors to
further work.




Community ecology theory often focuses on
questions of import primarily to communities of plants and animals,
examining models of interactions among a relatively small
number of species, whose traits are stably defined,
to explain patterns of coexistence and diversity.
In microbial ecology, where organisms of different taxa share
and exchange genes, and communities can be very diverse and
variable in composition over time and space,
theoretical questions particular to microbial ecology may be posed,
potentially driving ecological theory into new and productive
arenas.

\section{Acknowledgements}


This study was partially supported by a Models of Infectious Disease Agent Study
(MIDAS) grant from the US NIH/NIGMS to the University of California, San
Francisco (U01GM087728).
LW is grateful to Peter Ralph for a comment that motivated
this project, and to PR, Todd Parsons, Travis Porco,
Sarah Ackley, Rae Wannier, and several anonymous reviewers
for helpful conversation and feedback.


\bibliographystyle{vancouver}



\appendix
\section{Appendix}

\subsection{Analysis of conditions for conservation of trait structure across composition}
\label{app:analysis}

This section presents a mathematical analysis of the conditions
under which the model presented above conserves community
functional structure across variation in the selection of
species composing the community.

Let us assume $n_r$ resources, and
a community defined by $n_s$ species coexisting at
equilibrium, parametrized by constants $A_{ij}$,
$c_{ij}$, and $m_i$, and the functions $f_{ij}(\mathbf{R})$.
We wish to analyze conditions for the
vectors of trait abundances $\mathbf{T}$ and $\mathbf{E}$
at equilibrium
to be independent of the community composition.
That is, imagine a large pool of species described by
${A}$, ${c}$, and $m$ values and response functions ${f}$:
how must those values be constrained such that when a community
is assembled from a subset of those species,
the resulting trait abundances at equilibrium are the same
regardless
of which assemblage of those species was chosen.

This section will analyze the model equations in matrix form.
A model community is parametrized by constant matrices
$\mathbf{A}$ and $\mathbf{c}$ and vectors
$\mathbf{m}$ and $\mathbf{s}$, and a matrix
$\mathbf{f}$ whose entries are functions $f_{ij}$
of the resource abundances $R_j$.
The state of the model system is vectors
$\mathbf{X}$ and $\mathbf{R}$, and
derived vectors $\mathbf{T}$ and $\mathbf{E}$.
Many of these objects have equilibrium values
$\mathbf{X}^*$, $\mathbf{R}^*$, etc.
$\mathbf{1}$ is a vector of ones.
The operator $\odot$ stands for elementwise multiplication.


The relevant relations are as follows.
First there are the equilibrium conditions for
the $X$ variables,
\begin{dgroup*}
\begin{dmath}[number=eq-X] \label{eqX}
 (\mathbf{c} \odot \mathbf{A} \odot\mathbf{f}) \mathbf{1} = \mathbf{m}
\end{dmath},
\end{dgroup*}
and for the $R$ variables,
\begin{dgroup*}
\begin{dmath}[number=eq-R] \label{eqR}
 \mathbf{s} = (\mathbf{A} \odot \mathbf{f})^T \mathbf{X}
\end{dmath}.
\end{dgroup*}

The trait abundances are defined in terms of the $X$ variables:
\begin{dgroup*}
\begin{dmath}[number=T] \label{Tdef}
 \mathbf{T} = \mathbf{A}^T \mathbf{X}
\end{dmath}
\begin{dmath}[number=E] \label{Edef}
 \mathbf{E} = (\mathbf{A} \odot \mathbf{f})^T \mathbf{X}
\end{dmath}.
\end{dgroup*}

It follows immediately from the equilibrium condition (\ref{eqR})
that the equilibrium vector $\mathbf{E}^*$ of trait expression rates
is entirely determined by the vector of supply rates $\mathbf{s}$,
regardless of the population abundances $\mathbf{X}$ and
per-consumer uptake rates $\mathbf{f}$ that are the components
of the expression rates.
Because these expression rates are defined equal to the rate of uptake
of the resources, and equilibrium requires the rate of uptake to
equal the rate of supply, no specific conditions on the structure
of the community are needed to guarantee this result.
For this reason, trait expression rates in this model are conserved
across community structures.
However, conditions for conservation of the vector $\mathbf{T}^*$ of
rates of presence of traits are more restrictive and require more analysis.

The first equation (\ref{eqX}) is solved by finding values of $f_{ij}$
that bring the two sides to equality.
As written this equation is underdetermined as there are $n_s$
conditions for $n_sn_r$ values $f_{ij}$.
However, the $f$ variables are also constrained by their
dependence on the $n_r$-dimensional vector $\mathbf{R}$.
Because of that condition, the matrix $\mathbf{f}$
does not range freely over
$n_sn_r$ dimensions, but over an $n_r$-dimensional submanifold of
that space defined by the parametrization $\mathbf{f}(\mathbf{R})$:
\begin{dgroup*}
\begin{dmath}[number=f-R] \label{fR}
 \mathbf{f} = \mathbf{f}(\mathbf{R})
\end{dmath}.
\end{dgroup*}
An equilibrium matrix of resource uptake rates $\mathbf{f}^*$
is found by solving (\ref{eqX}) and (\ref{fR}) simultaneously.
The functional forms of the response functions $f_{ij}(\mathbf{R})$
can be substituted into (\ref{eqX}) to yield a system of $n_s$
equations in the $n_r$ variables $R_j$.
In the generic case, when $n_s=n_r$, this determines a unique solution
vector $\mathbf{R}^*$,
which determines the values of all entries of
the matrix $\mathbf{f}^*=\mathbf{f}(\mathbf{R}^*)$.
In other cases, multiple solutions for $\mathbf{R}^*$
and $\mathbf{f}^*$ may be possible.

Given $\mathbf{f}^*$, equilibrium population sizes are described
by (\ref{eqR}).
If the matrix $(\mathbf{A}\odot\mathbf{f}^*)^T$
is square and singular, then the vector $\mathbf{X}^*$
of equilibrium population sizes is the unique solution of (\ref{eqR}).
If the matrix is nonsingular, then there can be a space of solutions
for $\mathbf{X}^*$.

The trait abundances $\mathbf{T}^*$ must satisfy (\ref{Tdef})
given equilibrium values of $\mathbf{X}$.

Now let us imagine that the community's equilibrium trait abundances
can be predicted from the resource supply alone, without dependence on the
parameters describing the species in the community.
The above relations show that given a community structure,
both $\mathbf{T}^*$ and $\mathbf{s}$ 
are linearly related to $\mathbf{X}^*$.
For their relation to be independent of the community, let us assume
\begin{dgroup*}
\begin{dmath*}
\mathbf{T}^* = \mathbf{K} \mathbf{s}
\end{dmath*}
\end{dgroup*}
for some constant matrix $\mathbf{K}$.

This implies that
\begin{dgroup*}
\begin{dmath*}
\mathbf{T}^* = \mathbf{K} (\mathbf{A}\odot\mathbf{f}^*)^T \mathbf{X}^*
\end{dmath*}.
\end{dgroup*}
Comparing this to (\ref{Tdef}), it can be satisfied if
\begin{dgroup*}
\begin{dmath*}
\mathbf{K} (\mathbf{A}\odot\mathbf{f}^*)^T  = \mathbf{A}^T
\end{dmath*},
\intertext{or}
\begin{dmath*}
\sum_k {A}_{ik}{f}_{ik}(\mathbf{R}^*) {K}_{jk}  = {A}_{ij}
\end{dmath*}
\end{dgroup*}
for each $i$ and $j$.

Given a community parametrized by the constant matrix $\mathbf{A}$
and the functional forms $f_{ij}()$, the above equation describes
a set of $n_sn_r$ constraints on the resource concentrations $R^*_j$
and trait assignments $A_{ij}$,
which must be satisfied at equilibrium simultaneously with the previously
discussed constraints.

The above solves a general case of the problem, in which the entire
vector $\mathbf{T}^*$ of trait abundances is determined by the full vector 
$\mathbf{s}$ of supply rates.
The more restrictive case that each trait abundance $T^*_j$ depends only 
on the supply of resource $j$, rather than on all the resources'
supply rates, requires the matrix $\mathbf{K}$
to be diagonal.
In this case, the condition becomes 
\begin{dgroup*}
\begin{dmath*}
  A_{ij} f_{ij}(\mathbf{R}^*) k_{j} = A_{ij}
\end{dmath*}
\end{dgroup*}
for each $i$ and $j$,
where $k_j$ is the $j$'th diagonal entry of $\mathbf{K}$,
or
\begin{dgroup*}
\begin{dmath}\label{T-diag}
  f_{ij}(\mathbf{R}^*) = 1/k_j
\end{dmath}
\end{dgroup*}
for all $i$ and $j$ for which $A_{ij}$ is nonzero.

This condition implies that for each resource,
the equilibrium uptake rates $f^*$ of that resource must
be equal across its consumer species,
and equal to a value that
is uniform across different community structures.


Given that, the resource concentrations are those implied by these
values of $f$, and the species abundances are a solution of
(\ref{eqR}).
Note that that species abundances can vary depending on $\mathbf{A}$.
Also, I note that this condition can permit more than $n_r$ species
to coexist on $n_r$ resources, as it makes them compatible in a non-generic
way.

Note also, however, that because the operations of pointwise multiplication
and division,
matrix multiplication, and matrix inversion are continuous in the
values of all matrix entries, the above results have the property
that if the above two conditions are nearly met, that is,
if the $f$ and $A$ entries are within a suitably small
distance $\varepsilon$ of values that satisfy the conditions exactly,
then the trait abundances will be close to values that are exactly
conserved.
In other words, the functional regularity in question is approximately
achieved when the conditions are nearly enough met.
In this approximate but not exact case, the model does behave generically
and its diversity can be expected to be limited by the number of resources.

If resources have multiple consumers apiece,
the result that for each resource $j$,
$f_{ij}(\mathbf{R}^*)=1/k_j$
across all consumers $i$ of resource $j$
does not require that the response function 
have a uniform form across consumers,
$f_{ij}(\mathbf{R})\equiv f_j(R_j)$,
but that is certainly
one way it can be achieved.

The example models in this paper are a special
case of this condition,
constructed by assigning some fixed number $p$ of
traits to each species, and setting all consumers' functional
response curves for each resource equal,
with $m_i \equiv m$ and $c_{ij}\equiv c$.
In this case, (\ref{eqX}) is satisfied by
$f^*_{ij}=m/pc$ for all $i$ and $j$,
which also satisfies condition (\ref{T-diag})
with $k_j=pc/m$.
Equilibrium resource abundances are $R^*_j = f_{ij}^{-1}(m/pc)$
for each $j$, for any consumer $i$,
which is well-defined given that the functions
$f_{ij}()$ are assumed independent of $i$.
Under these assumptions, the above results imply
that $T^*_j = p\hspace{.1em}c s_j/m$ for each $j$.


\end{document}